\documentclass[pop,graphicx,reprint]{revtex4-1}

\usepackage{graphicx}
\usepackage{siunitx}
\usepackage{array,booktabs}

\begin{document}


\title{Experimental Study of Ion Heating in Obliquely Merging Hypersonic Plasma Jets}


\author{Samuel J. Langendorf}
\email[]{samuel.langendorf@lanl.gov}
\affiliation{Los Alamos National Laboratory, Los Alamos, NM 87545}

\author{Kevin C. Yates}
\affiliation{Los Alamos National Laboratory, Los Alamos, NM 87545}

\author{Scott C. Hsu}
\email[]{scotthsu@lanl.gov}
\affiliation{Los Alamos National Laboratory, Los Alamos, NM 87545}

\author{Carsten Thoma}
\affiliation{Voss Scientific, Albuquerque, NM 87108}

\author{Mark Gilmore}
\affiliation{University of New Mexico, Albuquerque, NM 87131}


\date{\today}

\begin{abstract}
In this experiment, we measure ion temperature evolution of collisional plasma shocks and colliding supersonic plasma flows across a range of species (Ar, Kr, Xe, N), Mach numbers, and collisionalities. Shocks are formed via the collision of discrete plasma jets relevant to plasma-jet-driven magneto-inertial fusion (PJMIF). We observe nearly classical ion shock heating and ion-electron equilibration, with peak temperatures attained consistent with collisional shock heating. We also observe cases where this heating occurs in a smooth merged structure with reduced density gradients due to significant intepenetration of the plasma jets. In application to PJMIF liners, we find that Mach number degradation due to ion shock heating will likely not be significant at the typical full-scale conditions proposed, and that a degree of interpenetration may be an attractive condition for PJMIF and similar approaches which seek to form uniform merged structures from discrete supersonic plasma jets.
\end{abstract}

\pacs{52.35.Tc, 52.30.-q, 52.30.Ex}

\maketitle 

\section{Introduction}
\label{sec:intro}
A shock wave is a dramatic and ubiquitous mechanism by which the bulk kinetic energy of a supersonic flow is converted to other forms, which may include thermal energy, energy of magnetic fields, and energy in accelerated particles. Shocks that occur via plasma collective effects over length scales much shorter than the post-shock collisional mean free path, termed collisionless shocks, are an exciting frontier area of study in plasma physics, with proposed relevance to the abundance of high energy particles and cosmic rays observed in the universe. Collisional plasma shocks, occurring by way of Coulomb collisions of the plasma charged particles, are perhaps less exotic than their collisionless counterparts but still contain rich physics. The structure of collisional plasma shocks has been studied analytically \cite{jaffrin1964structure} and a number of features have been predicted, including ion heating in excess of the local electron temperature, ion temperature relaxation via ion-electron collisional equilibration, and the formation of self-consistent electric fields within the shock thickness. In addition, when the energy density attained in the shocked region is high, radiative effects can become strong, plasma ionization and electrons can strongly effect transport, and a rich interplay of hot plasma physics and supersonic hydrodynamics can ensue. In recent years, with the increased development of high-energy-density laboratory plasma facilities, some of the these features of the collisional plasma shock can now be more readily observed in the laboratory \cite{swadling2013oblique, rinderknecht2018highly, young2019observation, melean2018design}. It is interesting to ask the question if our theories of collisional plasma shocks, having existed with minor modification for decades, are fully accurate and adequate to explain the range of experiments that are now possible.

Our motivation and approach in studying collisional plasma shocks is framed by research toward the development of hypersonic spherically imploding plasma liners as a novel driver architecture for the compression and heating of magneto-inertial fusion \cite{lindemuth83,kirkpatrick95,thio08,lindemuth09,lindemuth15} targets. This architecture has been termed plasma-jet driven magneto-inertial fusion, or PJMIF \cite{thio99, hsu09, hsu2012spherically}, and aims to form such a spherically imploding liner via the merging of a spherical array of discrete supersonic plasma jets. An important metric for PJMIF liner performance is the time-averaged liner Mach number, which must remain high for the liner to effectively compress a target at a substantial standoff distance. If collisional plasma shocks form when the discrete plasma jets merge, this could elevate the liner ion temperature, increase sound speed, and degrade the liner Mach number, resulting in decreased performance of the implosion \cite{awe11,langendorf2017semi}. Plasma shocks may also cause non-uniformities in the liner density, which may seed hydrodynamic instabilies in the later implosion phases. To experimentally investigate these flows, plasma jet merging experiments have been conducted on the Plasma Liner Experiment (PLX) at Los Alamos National Laboratory over the past several years \cite{merritt13,merritt14,moser2015experimental,adams2015observation,hsu2018experiment}. Recently, we have performed experiments to measure the ion heating and temperature evolution during the jet merging and collisional plasma shock formation \cite{langendorf2018experimental}, which is the focus of this paper, and to benchmark computer models of the jet merging with six and seven jets, which will be discussed in a separate publication \cite{yates19personal}.

The structure of this paper is the following: in Section~\ref{sec:setup}, an overview of the experimental setup is given, Section~\ref{sec:reduc}, procedures of diagnostic data reduction are described, useful to inform the primary results of ion temperature evolution, and in Section~\ref{sec:results} these results are presented and discussed. In Section~\ref{sec:liner}, implications of the current results for the formation of PJMIF-scale plasma liners are stated and the significance of Mach number degradation due to shock ion heating is assessed.

\section{Experimental Setup}
\label{sec:setup}
The PLX facility consists of a 9-\si{ft} diameter spherical vacuum chamber that is equipped with six coaxial plasma guns, which can be fired in any permutation to produce supersonic plasma jets that can be merged/collided at the different angles corresponding to the ports on the chamber. In the current experiments, 2 guns are fired at a time producing plasma jets with nominal number densities ~ $10^{16}~\textrm{cm}^{-3}$, temperatures $\sim$ 1.5 eV, and velocities $\sim$ 50 \si{km/s}. 

Plasmas are formed in the guns via a gas-puff injection and preionization current pulse, and are accelerated by a primary current discharge between coaxial electrodes. While there initially exists a strong magnetic field associated with the drive current, the plasma jets formed in these experiments are dense and cool enough ($\sim 10^{16}$ \si{cm^{-3}}, $\sim$ few eV) that the characteristic time for the magnetic field to decay is much less than the travel time before the jets collide. Hence in this experiment we assume the jets to be unmagnetized.

\begin{figure}[htbp]
\includegraphics[width=3.37truein]{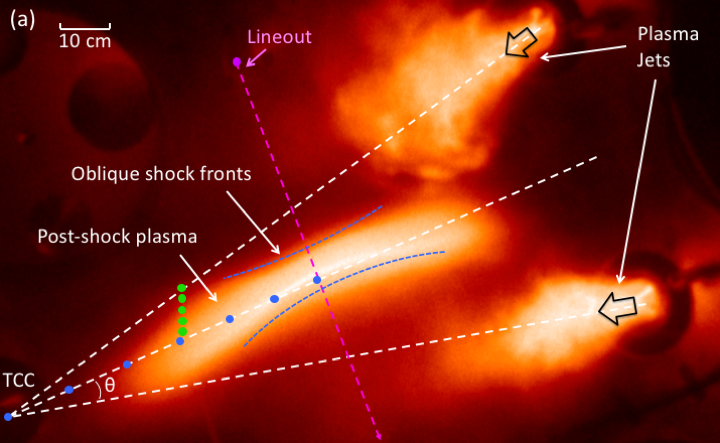}
\includegraphics[width=3.37truein]{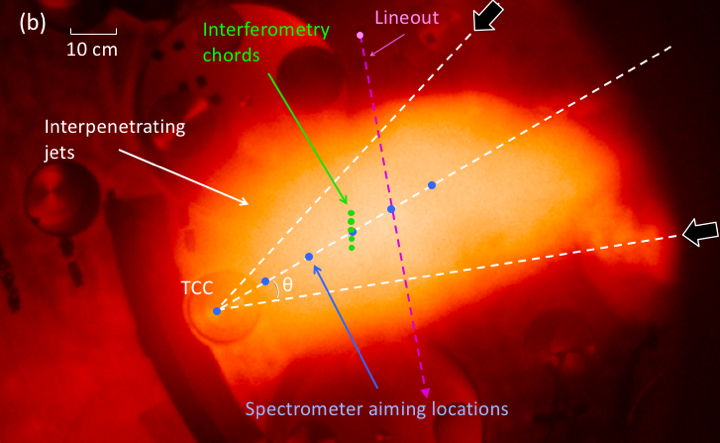}
\caption{Fast framing camera images of argon plasma jets (10 \si{ns} exposure, false color, log of intensity) merging obliquely at (a) 11.6\si{\degree} and (b) 20.5\si{\degree} merging half-angle. Spatial locations of spectrometer views and interferometry chords, and lineout of Fig.~\ref{fig:lineouts} overlaid.}
\label{fig:setup}
\end{figure}

 Jet merging experiments are performed at different merging half-angles of $11.6\si{\degree}$ and $20.5\si{\degree}$, with diagnostic data collected at multiple times and spatial positions across repeated shots. Figure~\ref{fig:setup} shows framing camera images of merging plasma jets at the different angles, with diagnostic locations overlaid. Experiments at both angles are repeated for each of four different gas species, Ar, Kr, Xe, and N. A further overview of the PLX facility in roughly the configuration used for these experiments is available in Ref.~\citenum{hsu2018experiment}, the reader is referred there for additional information, while we will describe here the essential features and details of the current experiments.

\section{Data Reduction}
\label{sec:reduc}
Parameters important to the jet merging are the ion-ion slowing length, ion-ion mean free path, and the relative velocity / Mach number in the shock direction. These parameters depend on the plasma density, electron and ion temperatures, velocity, and ionization state, which are inferred from multiple diagnostics. A summary of the results is given in Table~\ref{tab:conditions}, and more detail of the separate measurements is given in the following sections. 

\subsection{Plasma Jet Velocity}
\label{sec:velocity}
Plasma jet velocity is measured by analysis of the time evolution of the light emission of the plasma jets as they are fired past a pair of pinholed plastic optical fibers looking perpendicular to each gun. Light is delivered via the fibers to silicon photodiodes and digitized at 100 \si{MS/s}. The separation of each photodiode pair is 2 \si{cm}, which is sufficiently short to lead to unambiguous correlation of the data traces in the majority of cases. By analysis of the time separation of the two photodiode traces, the jet velocity is deduced. 

\begin{figure}[htbp]
\includegraphics[width=3.37truein]{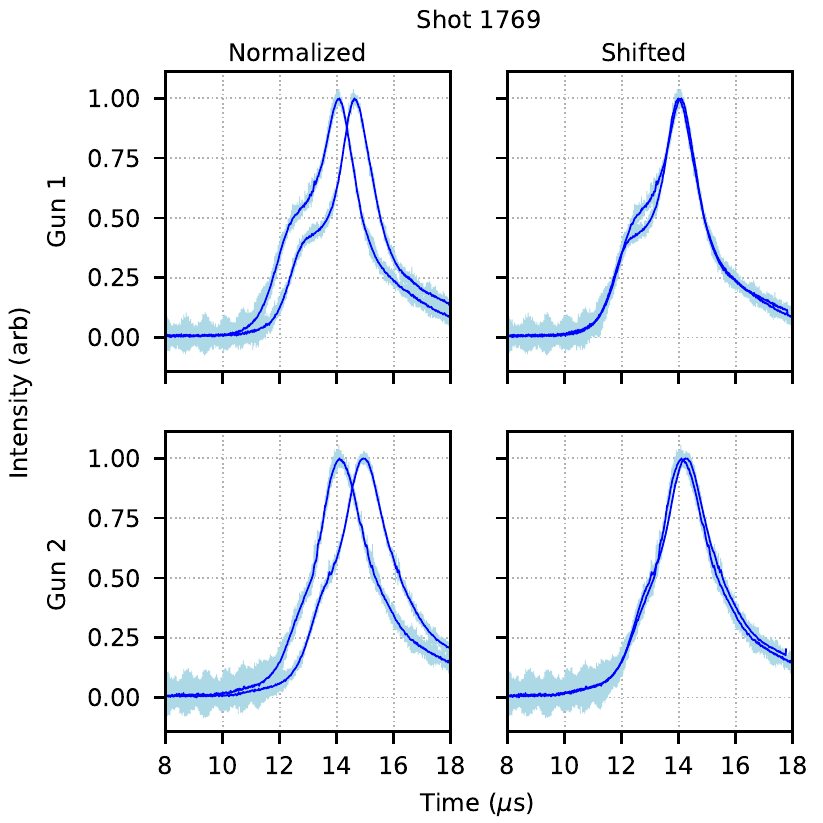}
\caption{Example of normalized raw, smoothed, and and time-shifted photodiode data for deduction of jet velocity of shot 1769. Time shift was found based on median time difference between an array of levels spanning 10\% - 90\% of the leading edge.}
\label{fig:pd}
\end{figure}

Multiple methods were tested and compared for automatic processing of the photodiode signals to determine the jet velocity. In all cases, the signals were normalized to a common amplitude scale and smoothed by moving average with a window of four samples, which drastically attenuated a systematic periodic noise present in the raw data. In the first method, the leading signal was time-shifted by an iterated amount and the correlation with the second signal was computed. The time shift corresponding to the jet velocity was determined as the shift having resulted in maximum correlation of the two signals. In the second method, the time differences of first crossings of a series of levels between 10\% and 90\% of the maximum signal were computed, and the time shift was determined as the median of these values. For well-behaved signals these methods agreed closely, however across the breadth of signals collected the second method proved more robust as assessed by visible inspection, and was thus used for determination of the jet velocity. An example of the recorded, smoothed, and time-shifted data (using the second method) is shown in Figure~\ref{fig:pd}.

\subsection{Post-merge Density}
\label{sec:density}
\begin{figure*}[htbp]
\includegraphics[width=3.37truein]{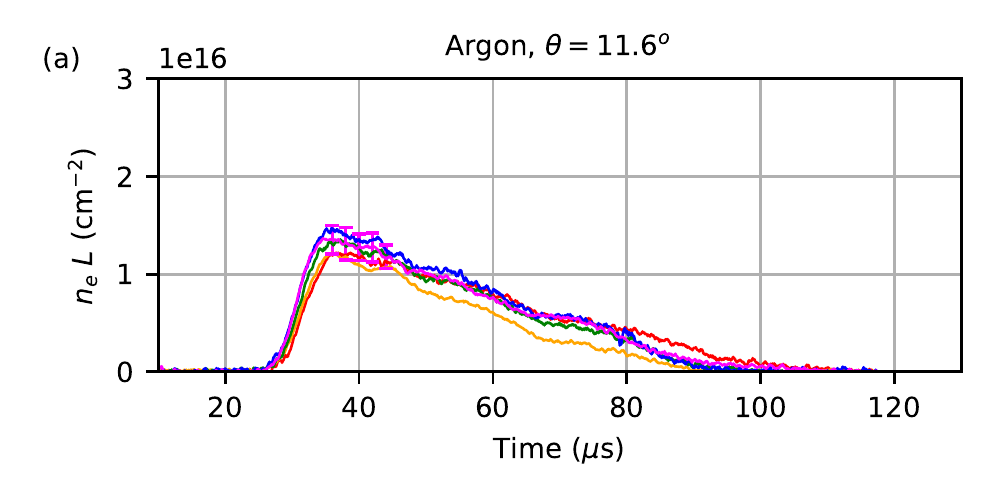}
\includegraphics[width=3.37truein]{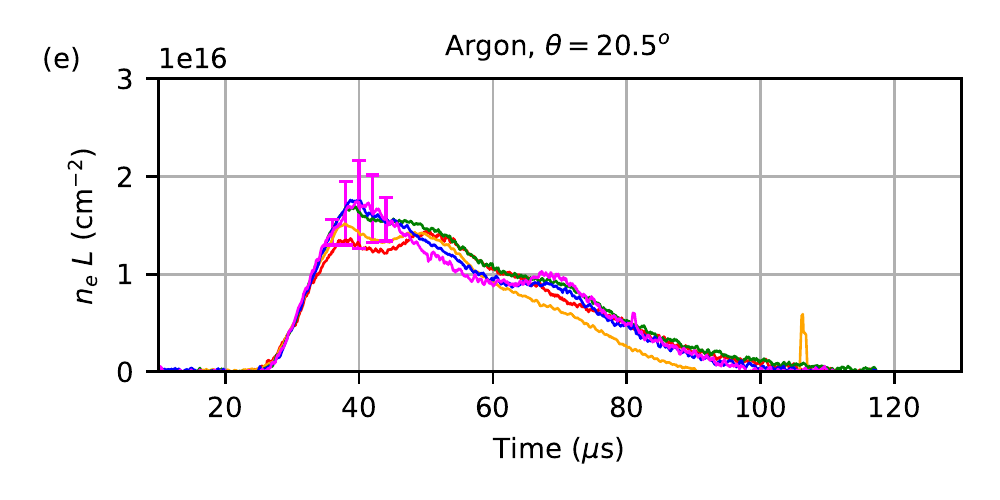}
\includegraphics[width=3.37truein]{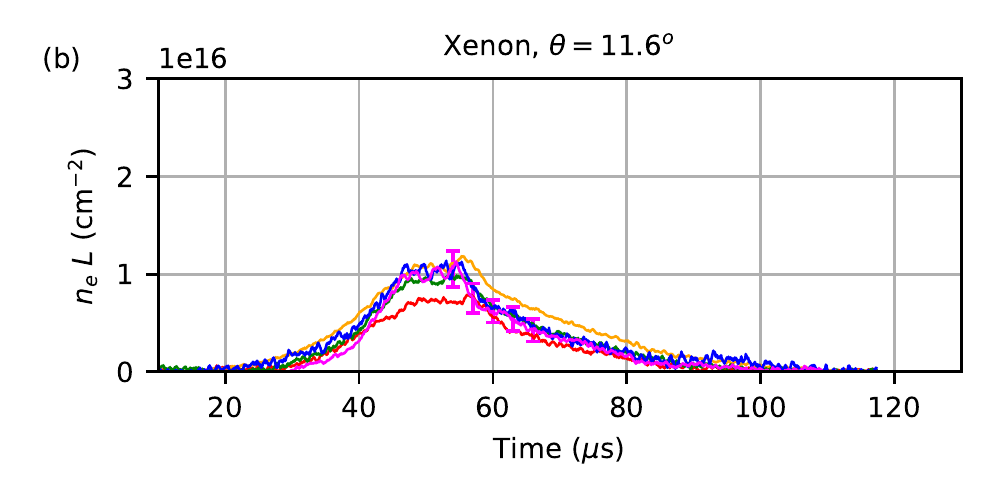}
\includegraphics[width=3.37truein]{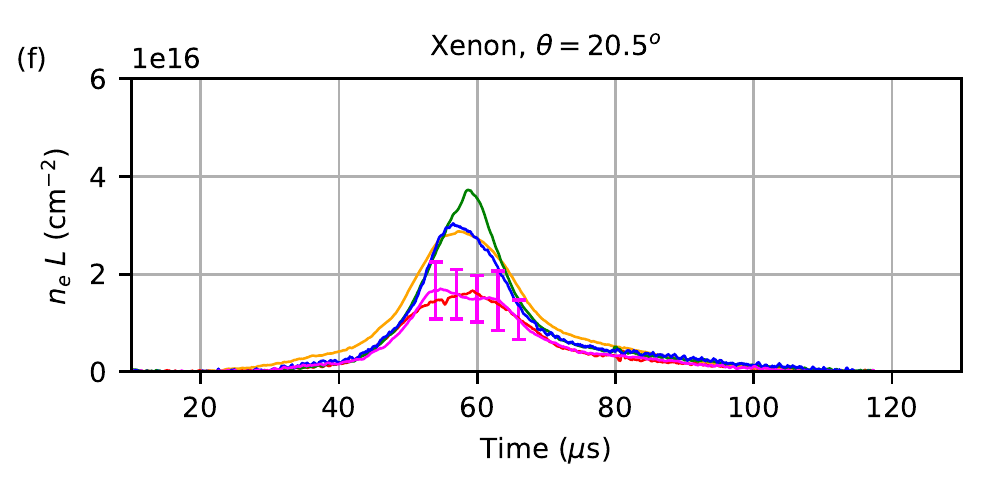}
\includegraphics[width=3.37truein]{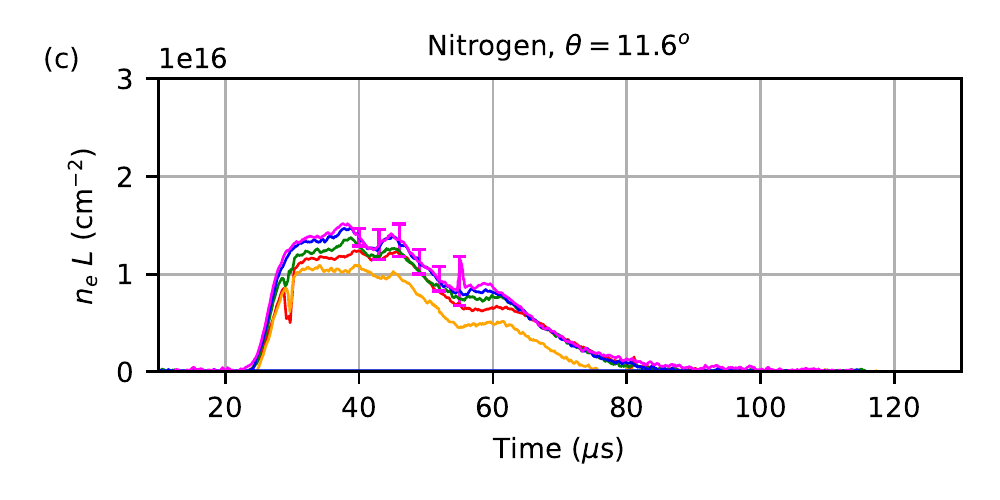}
\includegraphics[width=3.37truein]{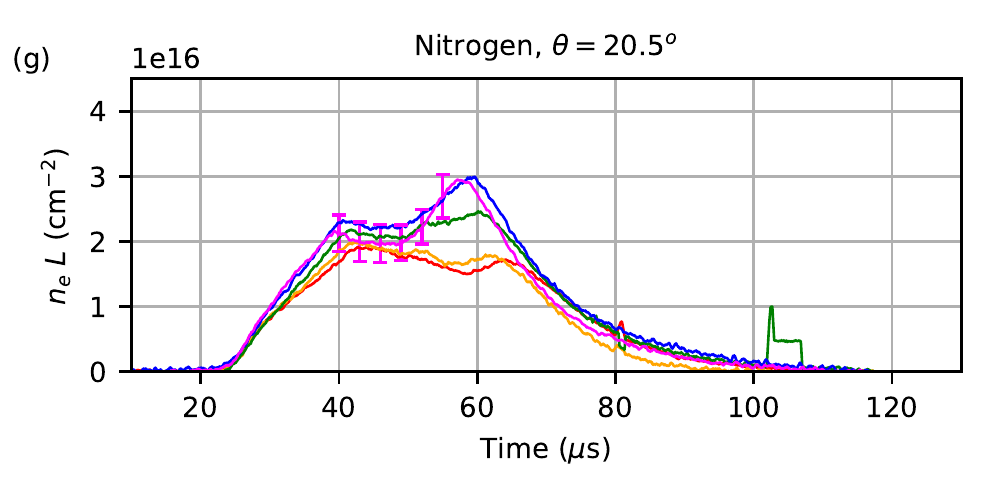}
\includegraphics[width=3.37truein]{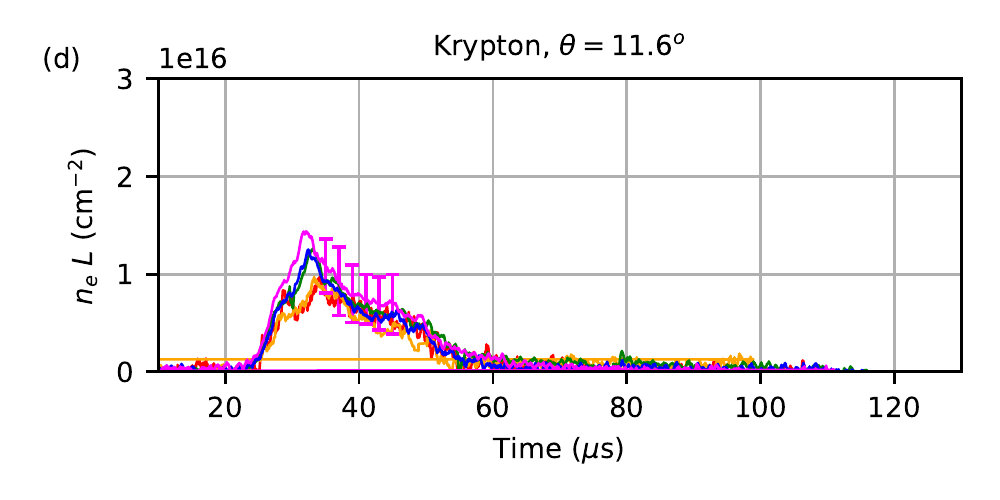}
\includegraphics[width=3.37truein]{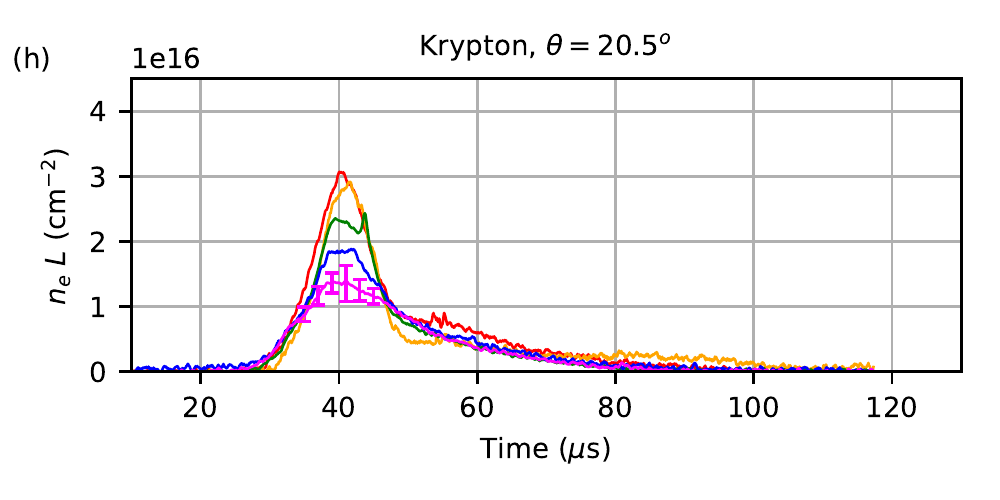}
\caption{Interferometer measurements of merged plasma line-integrated electron density, averaged over ten shots. Lines indicate time-resolved data from 5 transverse measurement chords, 1.5 cm spacing between chords.  Errorbars (placed sparsely for legibility) indicate $\pm 1\sigma$ range of shot-to-shot variation.}
\label{fig:interferom}
\end{figure*}

Line-integrated electron density of the post-merge / post-shock plasma is measured using a multi-chord laser interferometer~\cite{merritt2012multi}. For these experiments, five chords of the interferometer are positioned transverse to the axis of symmetry of the 6 jet mounting pattern on the chamber. Interferometer data are recorded at 40 \si{MS/s} and smoothed by application of a 20-sample moving average. The interferometer phase shifts observed are solely in the positive direction, so we do not see evidence as was seen in previous work of large preponderance of neutral species in the jet trailing regions. This may be due to differences in the gun designs used between the two experiments, and also due to increased damping of the pulsed power system present in these experiments decreasing the production of colder trailing jets. An average post-shock electron density is inferred using estimates of the plasma length from lineouts of visible framing camera images. Averaged interferometer data traces are shown in Figure~\ref{fig:interferom} for the varied gas species and merging angles. These data are then combined with path length estimates from framing camera images to constrain the post-merge electron number density.

One additional note from the time-resolved interferometer measurements is the observation of the length of the jet after it has been traveled most of the way to the center of the vacuum chamber.  It is seen that xenon has similar or shorter overall jet length than the other species, despite being fired at a considerably lower velocity. This is due to the heavier atomic weight of the xenon atoms increasing the overall Mach number of the jet, and spreading occurring more slowly at the decreased sound speed. An important challenge for standoff plasma liners is to prevent expansion of the liner en route to engagement with the target as has been shown in previous work \cite{langendorf2017semi} -- these results highlight the fact that using the heaviest species possible for the liner is beneficial in that regard. 

\subsection{Post-merge Electron Temperature and Ionization State}
\label{sec:temp}
\begin{figure}[htbp]
\includegraphics[width=3.3truein]{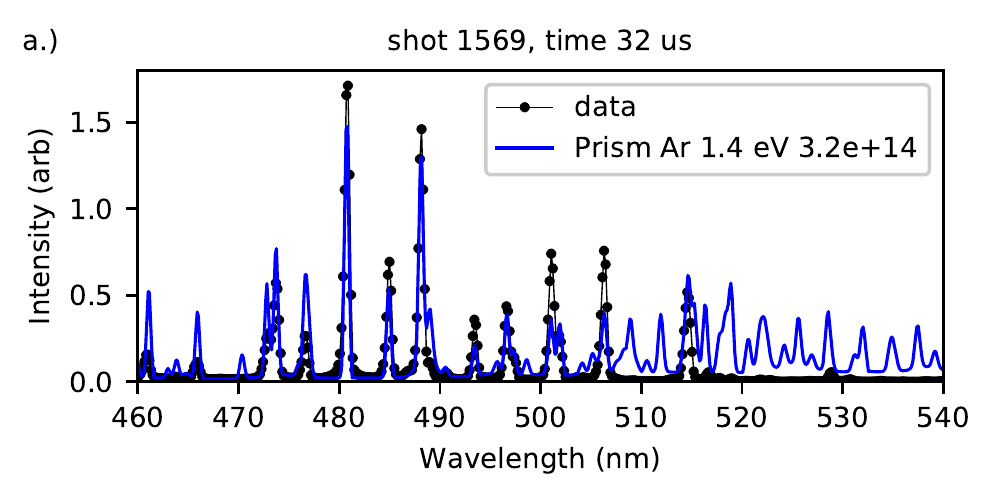}
\includegraphics[width=3.3truein]{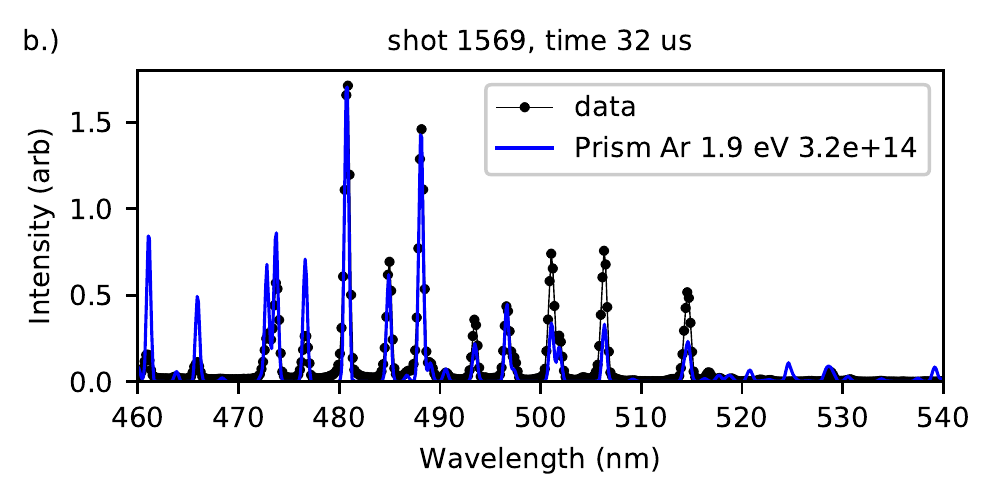}
\includegraphics[width=3.3truein]{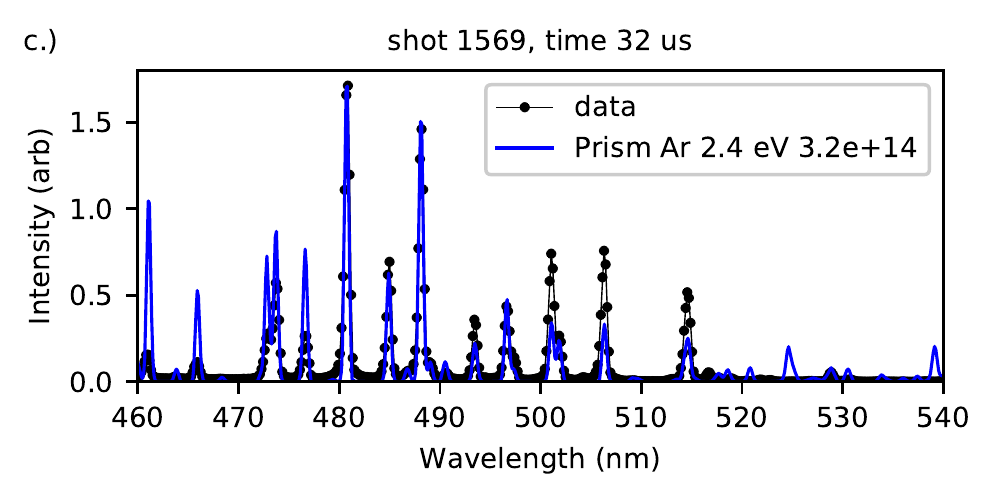}
\caption{Survey spectrometer data from shot 1569, compared to PrismSPECT simulations with differing plasma electron temperatures. The temperature is bounded by presence or absence of predicted spectral lines.}
\label{fig:spectro}
\end{figure}

The electron temperature and ionization state of the post merge plasma is constrained by analysis of broadband line-integrated visible emission spectra and comparison with steady-state NLTE atomic physics modeling performed using the PrismSPECT software package \cite{macfarlane2003prismspect}. Primarily, emission lines corresponding to singly-ionized species of the plasma jet species are observed. Ultimately, this analysis leads to the conclusion that post-merge electron temperatures remain in the range of a few eV and average ionization states are not far in excess of unity. Bounds for these plasma parameters are determined by manually demarcating the region of best fit. An example of spectrometer data fitting is shown in Figure~\ref{fig:spectro}, and overall determined bounds are listed in Table \ref{tab:conditions}.

\begin{table*}[!t]
\caption{Summary of experimental parameters. The $n_e$, $T_i$, $T_e$, $\bar{Z}$,
and ion--ion mean free path $\lambda_i$ are average, post-merge values.   
The jet--jet interpenetration length $L_{ii,s}$ [see Eq.~(\ref{eqn:ionslowing})], 
counter-streaming speed $v_{cs}=2v_{jet}\sin\theta$,
and jet counter-streaming Mach number $M_{cs}=v_{cs}/[\gamma k (T_i + \bar{Z}T_e)/m_i]^{1/2}$
are average, pre-merge values.  The average $L_{ii,s}$ and $\lambda_i$
values are not intended to be precise but to
provide insight into the collisionality regime.  The error ranges for $v_{jet}$, $v_{cs}$, $n_e$, and $T_i$
are $\pm 1\sigma$ of the variation over multiple shots; those for $T_e$ and $\bar{Z}$
represent uncertainties based on comparisons with PrismSPECT spectral modeling.
\label{tab:conditions}}
\newcolumntype{C}{ @{}>{${}}c<{{}$}@{} }
\begin{ruledtabular}
\begingroup
\setlength{\tabcolsep}{6pt} 
\begin{tabular}{c|*{8}{>{$}r<{$}@{}>{$}l<{$}}>{$}l<{$}}
  Case&\multicolumn{2}{c}{(a)}&\multicolumn{2}{c}{(b)}&\multicolumn{2}{c}{(c)}&\multicolumn{2}{c}{(d)}&\multicolumn{2}{c}{(e)}&\multicolumn{2}{c}{(f)}&\multicolumn{2}{c}{(g)}&\multicolumn{2}{c}{(h)}\\
\hline
  Half-angle $\theta$ &\multicolumn{2}{c}{11.6\si{\degree}}&\multicolumn{2}{c}{11.6\si{\degree}}&\multicolumn{2}{c}{11.6\si{\degree}}&\multicolumn{2}{c}{11.6\si{\degree}}&\multicolumn{2}{c}{20.5$\si{\degree}
  $}&\multicolumn{2}{c}{20.5\si{\degree}}&\multicolumn{2}{c}{20.5\si{\degree}}&\multicolumn{2}{c}{20.5\si{\degree}} \\
  Species&\multicolumn{2}{c}{Ar}&\multicolumn{2}{c}{Xe}&\multicolumn{2}{c}{N}&\multicolumn{2}{c}{Kr}&\multicolumn{2}{c}{Ar}&\multicolumn{2}{c}{Xe}&\multicolumn{2}{c}{N}&\multicolumn{2}{c}{Kr} \\
$v_{\mathrm{jet}}$ (\si{km/s})&41.5&\ \pm\ 4.5&24.3&\ \pm\ 3.1&44.8&\ \pm\ 4.6&64.8&\ \pm\ 18.1&42.1&\ \pm\ 4.8&27.4&\ \pm\ 3.6&52.2&\ \pm\ 3.5&57&\ \pm\ 7.5\\
$v_{cs}$ (\si{km/s})&16.7&\ \pm\ 1.8&9.8&\ \pm\ 1.2&18.1&\ \pm\ 1.9&26.1&\ \pm\ 7.3&29.4&\ \pm\ 3.3&19.2&\ \pm\ 2.5&36.5&\ \pm\ 2.4&39.8&\ \pm\ 5.3\\
$n_{e}$ ($10^{14}$ cm$^{-3}$)&4.0&\ \pm\ 0.5&4.8&\ \pm\ 0.8&4.6&\ \pm\ 0.4&3.8&\ \pm\ 1.8&4.6&\ \pm\ 1.0&13&\ \pm\ 5.1&8.9&\ \pm\ 1.4&11.6&\ \pm\ 2.9\\
Peak $T_{i}$ (eV)&18.1&\ \pm\ 6.5&25.6&\ \pm\ 3.2&10.2&\ \pm\ 2.2&31.7&\ \pm\ 21.3&32.0&\ \pm\ 2.3&40.6&\ \pm\ 10.0&16.6&\ \pm\ 2.8&45.6&\ \pm\ 10.4\\
  $T_{e}$ (eV)&2.0&\ \pm\ 0.4&1.7&\ \pm\ 0.4&1.7&\ \pm\ 0.9&1.4&\ \pm\ 0.6&2.0&\ \pm\ 0.4&1.7&\ \pm\ 0.4&2.6&\ \pm\ 0.8&1.4&\ \pm\ 0.6\\
$\bar Z$ &1.0&\ \pm\ 0.1&1.2&\ \pm\ 0.2&1.0&\ \pm\ 0.2&1.0&\ \pm\ 0.2&1.0&\ \pm\ 0.1&1.2&\ \pm\ 0.2&1.1&\ \pm\ 0.2&1.0&\ \pm\ 0.2\\
$L_{ii,s}$ (cm)&\multicolumn{2}{c}{2.5}&\multicolumn{2}{c}{1.5}&\multicolumn{2}{c}{0.2}&\multicolumn{2}{c}{56.2}&\multicolumn{2}{c}{26.6}&\multicolumn{2}{c}{10.2}&\multicolumn{2}{c}{2.9}&\multicolumn{2}{c}{190} \\
$\lambda_{i}$ (cm)&\multicolumn{2}{c}{1.9}&\multicolumn{2}{c}{1.6}&\multicolumn{2}{c}{0.5}&\multicolumn{2}{c}{2}&\multicolumn{2}{c}{3.3}&\multicolumn{2}{c}{1.4}&\multicolumn{2}{c}{0.4}&\multicolumn{2}{c}{2.6} \\
$M_{cs}$&\multicolumn{2}{c}{4.2}&\multicolumn{2}{c}{4.7}&\multicolumn{2}{c}{2.9}&\multicolumn{2}{c}{11.4}&\multicolumn{2}{c}{7.4}&\multicolumn{2}{c}{9.1}&\multicolumn{2}{c}{4.6}&\multicolumn{2}{c}{17.3} \\
\end{tabular}
\endgroup
\end{ruledtabular}
\end{table*}

\subsection{Ion Temperature}


The primary focus of this work is to measure the level of ion heating that occurs in the jet merging and/or collisional plasma shock formation. This measurement is performed by analysis of the Doppler broadening of the plasma emission lines. It is important in this study to consider the possible effects of alternative sources of broadening, including Stark broadening and the effects of Doppler shift caused by the significant plasma jet bulk velocity. While Stark broadening is not expected to be significant in high Z species at these densities $\sim 10^{14}$~\si{cm^{-3}}, the supersonic jet bulk velocities are by definition significant. It was initially expected that the Doppler shift due to the bulk jet velocities would be helpful in providing an additional measure of the jet velocity -- in practice, the limited angles available for view produced signals that were difficult to interpret. The plasma was thus observed from a view that was as near to normal to the plane formed by the trajectories of the two guns as possible. The achieved view was estimated from the chamber mechanical design to be within $\sim 2\si{\degree}$ of normal.

Since the plasma jet does not remain perfectly collimated as it travels towards the chamber center, but rather expands, it is to be expected that some broadening will be observed due to peripheral bulk plasma motion towards and away from the spectrometer view. However the parts of the plasma in motion are likely to be the least dense and thus contribute only a small amount to the radiation observed in the line-integrated spectrometer view through the plasma, decreasing as the density squared. Thus, no correction is made to the observed data for the possibility of these bulk motions, from the reasoning that corrections to data should be avoided when it is not certain that they are justified.

\begin{figure}[htbp]
\includegraphics[width=3.3truein]{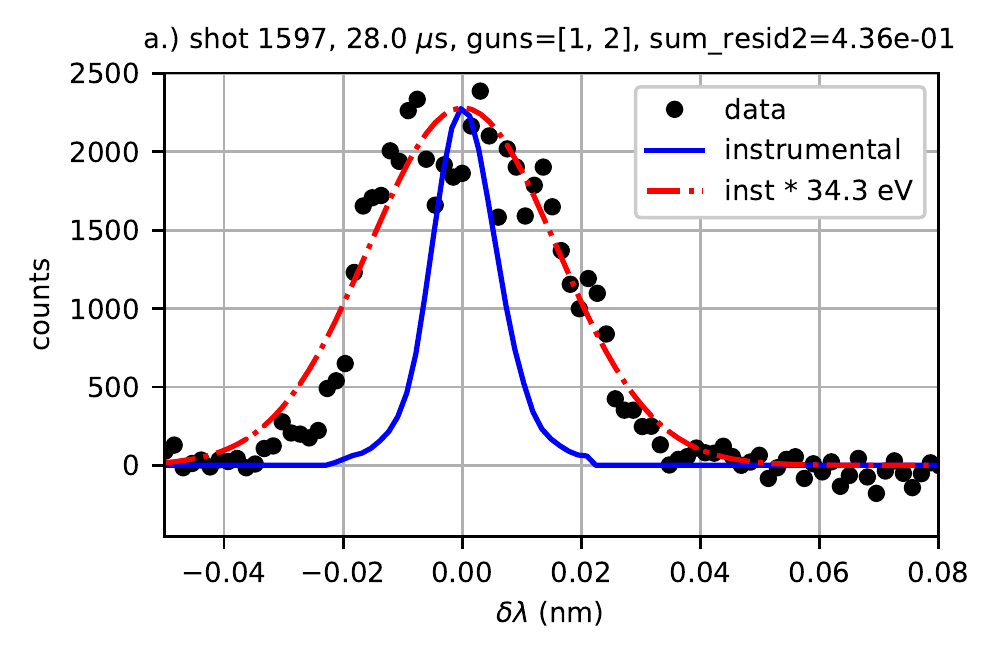}
\includegraphics[width=3.3truein]{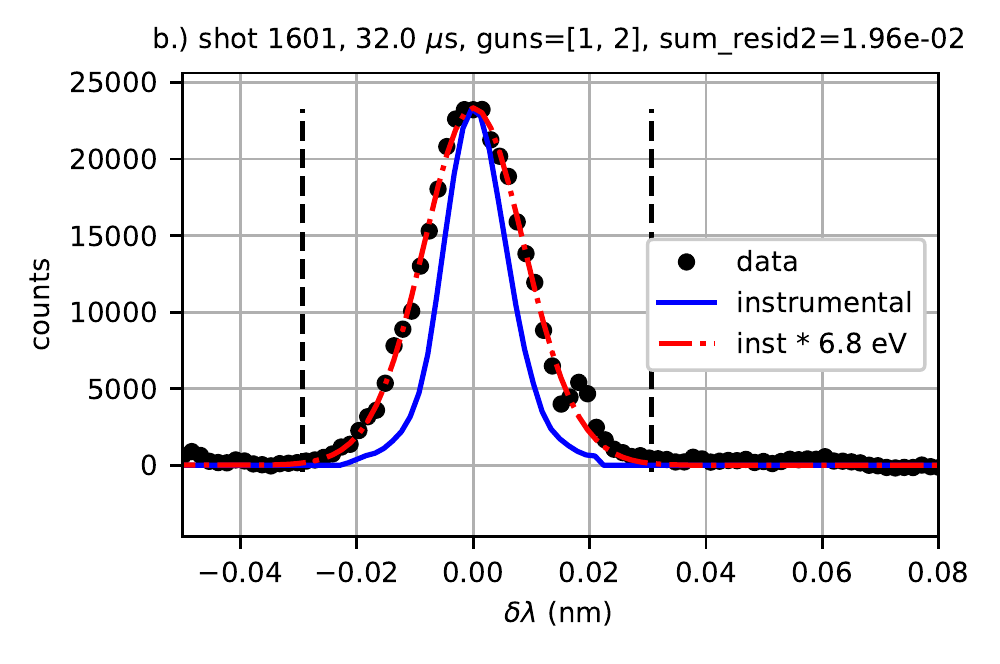}
\caption{Example raw Doppler broadening data. Data acquired at early times in the jet merger often showed non-Gaussian line shapes such as (a), while well into the jet collision curves tended to a Gaussian as shown in (b).}
\label{fig:doppler}
\end{figure}

Figure~\ref{fig:doppler} shows examples of recorded line shapes and forward-fit convolutions of Doppler-broadened line shape and instrumental broadening. Figure~\ref{fig:doppler}(a) is representative of line shapes often recorded earlier in time in the jet merging process, when flat-topped and non-gaussian shapes were often observed. Later in experimental time, shapes came to resemble Figure~\ref{fig:doppler}(b), where the convolution of instrumental broadening and Gaussian shape fit the data well. In generating the principle summary figures (Figs. \ref{fig:dopplerSpatial} - \ref{fig:dopplerTi}) from these data, a threshold of goodness-of-fit to the broadened Gaussian consisting of the sum of squares of the fitting residuals is employed, such that shapes such as Figure~\ref{fig:doppler}(a) are indicated or excluded -- such plots are thus focusing on the relatively well-behaved data points where the plasma has assumed a single temperature and absence of drastically divergent bulk motions is inferred. More discussion of the temporal and spatial variance of lineshapes and their physical interpretation is included in Section~\ref{sec:iontemp}.

\section{Results and Discussion}
\label{sec:results}

\subsection{Jet Structure}
\label{sec:structure}

A qualitative but nonetheless valuable look at the jet structure can be gained from analysis of framing camera images of the jet merging.  The majority of the images taken in the experiment were from an angled and incomplete view making it hard to discern details of the shock structure, but overall coarse insights can be gained from analysis of images such as those of Fig~\ref{fig:setup}. 
\begin{figure}[htbp]
\includegraphics[width=3.3truein]{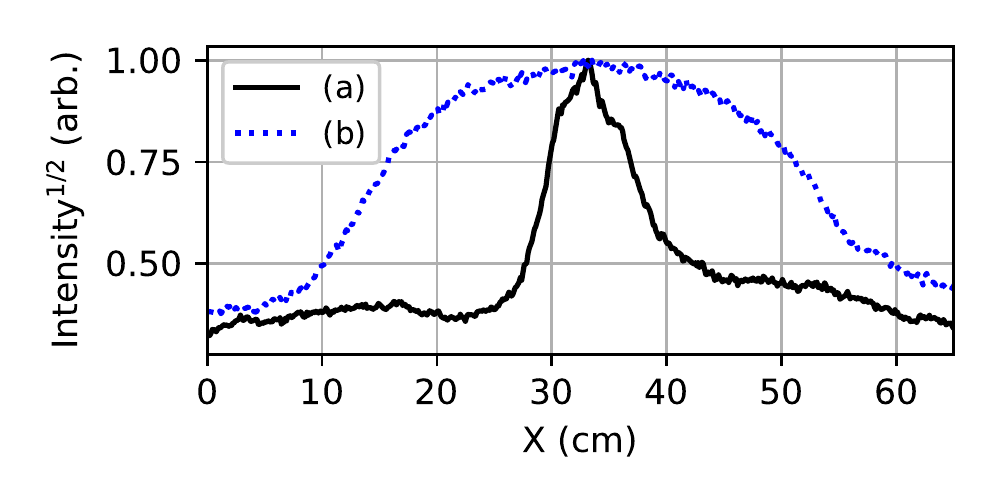}
\caption{Lineout of the square root of intensity of visible emission images of Fig~\ref{fig:setup} (a) and (b), illustrative of plasma density gradient length scales.}
\label{fig:lineouts}
\end{figure}
Fig.~\ref{fig:lineouts} shows a comparison between argon jet mergers with significantly varied relative normal velocity, mainly due to the change in merging half-angle from $11.6\si{\degree}$ to $20.5\si{\degree}$. We calculate the ion-ion slowing length as
\begin{equation}
L_{ii,s} = \frac{v_{cs}}{4 \nu_{ii,s}} = \frac{v_{cs}}{4} \left[\num{9e-8}n_i \bar Z^4\Lambda_{ii}  \left( \frac{2}{\mu}  \right) 
\frac{\mu^{1/2}}{\epsilon^{3/2}} \right]^{-1},
\label{eqn:ionslowing}
\end{equation}
in which $v_{cs}$ is the counter-streaming jet speed in the merging direction, $n_i$ is pre-merge ion number density, $\bar Z$ is the average ionization state, $\Lambda_{ii}$ is the ion-ion Coulomb logarithm in the case of counterstreaming fast ions, $\mu$ is the ion mass normalized by the proton mass, and $\epsilon$ is the ion directed energy associated with $v_{cs}$. The factor of $1/4$ results from integration over the stopping trajectory as the slowing rate varies with the particle speed \cite{messer2013nonlinear}. At the wider merging angle, the interpenetration distance is greatly increased due to the 4th power effect in equation (\ref{eqn:ionslowing}). The gradients observed in Fig.~\ref{fig:lineouts} shows that indeed the sharpness of the merging layer becomes greatly smoothed out when the interpenetration length is significantly greater than the mean free path. In this way, collisional plasma shock formation cannot be said to have occurred even though the merging velocities are supersonic, due to the finite-Knudsen-number effect of the ion interpenetration.

\subsection{Ion Temperature}
\label{sec:iontemp}

\begin{figure}[htbp]
\includegraphics[width=3.3truein]{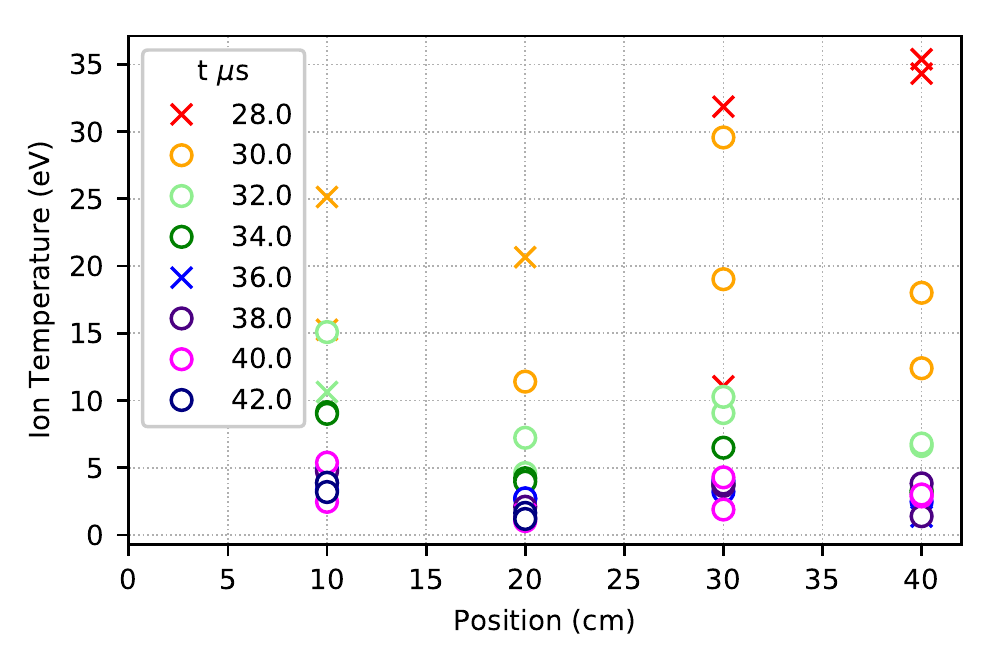}
\caption{Ion temperatures inferred by doppler broadening for argon jets colliding at $\theta = 11.6$\si{\degree}. Open symbols indicate measurements that were accepted by the goodness-of-fit criterion, x samples indicated measurements rejected.}
\label{fig:dopplerSpatial}
\end{figure}

Figure~\ref{fig:dopplerSpatial} shows ion temperatures inferred from Doppler broadening in the case of Argon jets colliding at the shallow (11.6\si{\degree}) merging half-angle, as a function of both spatial position and experimental time. Associated plasma parameters are given in Table~\ref{tab:conditions}(a), indicating that the merging should be fairly collisional and supersonic with mean free path and interpenetration length being of similar order, a few cm. It is also indicated which data points meet the goodness-of-fit threshold and which do not. It seen that at early times in the jet merging, line fits are often rejected and overall inferred temperatures are high. As time progresses, line fits become accurate and the ion temperature assumes decreased values.  

From the interferometry measurements, it is observed that the peak density occurs at 36 \si{\micro\second} at 20 \si{\centi\meter}. Given the velocity of the jet, it can be thus reasoned that the peak density crosses the spectrometer measurement locations at corresponding times, being approximately 34 \si{\micro\second} crossing the 30 \si{\centi\meter} location and 32 \si{\micro\second} crossing the 40 \si{\centi\meter} location.  It is thus clear that the early time data points in which flat-topped and non-gaussian shapes are observed correspond to the leading / early part of the jet merging, while the overall density is still quickly increasing.  These line shapes could thus be due to non-uniform bulk velocities if the leading edges of the jets are mutually interpenetrating and scattering.  More detailed characterization of the velocity distribution evolution on the initial merging is an attractive prospect for future work -- in the present data set we mainly resolve dynamics after the velocity has relaxed to a single Maxwellian, from approximately 32 \si{\micro\second} onward in this case.

\begin{figure}[htbp]
\includegraphics[width=3.3truein]{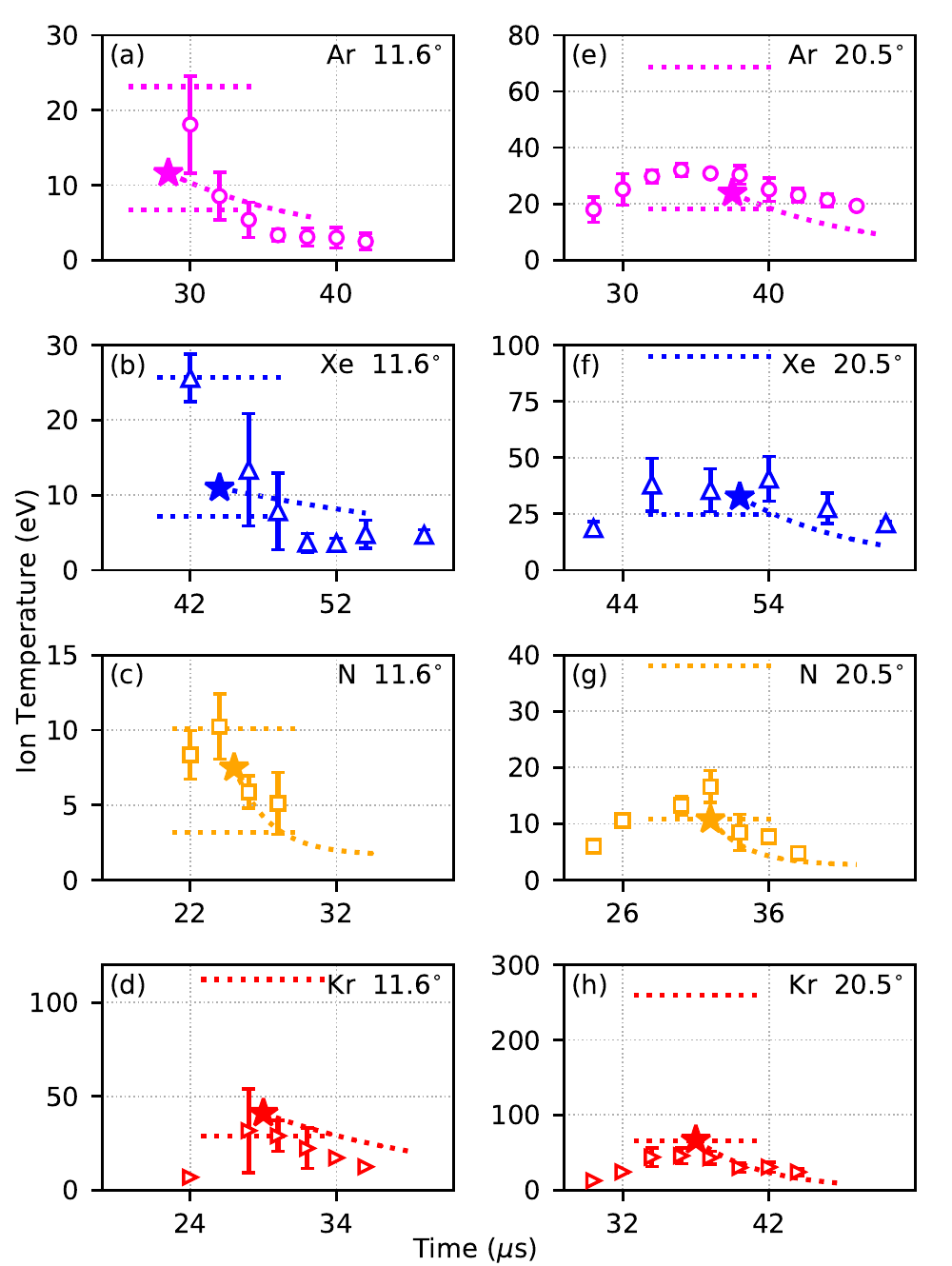}
\caption{Ion temperatures inferred by doppler broadening of plasma self-emission lines across breadth of collected data. Errorbars indicate standard deviation of shot to shot variation. Horizontal dotted lines indicated predicted peak temperatures using equation~(\ref{eqn:peakT}). Sloped dotted line indicates predicted ion-electron equilibration rate using equation~(\ref{eqn:coolingTime}). Stars indicate peak temperatures found in 1D multi-fluid CHICAGO simulations.}
\label{fig:dopplerTi}
\end{figure}

Inspection of Figure~\ref{fig:dopplerSpatial} reveals that the ion temperatures are more strongly dependent on time than spatial location, indicative of relatively global heating and cooling along the length of the oblique jet merger. Figure~\ref{fig:dopplerTi} shows the time evolution of measured post-shock ion temperatures inferred via Doppler broadening for the varied jet species and jet merging angles. The results are compared to eq.~(\ref{eqn:peakT}), a result for the peak ion heating in a two-fluid treatment (electron and ion fluids) of a collisional plasma shock, if it is assumed that all kinetic energy goes to ion heating (Ref.~\cite{langendorf2018experimental} supplemental material),
\begin{equation}
\frac{T_{i2}}{T_{i1}}  = \left[ 1 + \frac{2 (\gamma - 1)}{(\gamma + 1)^2} \frac{\gamma M^2 + 1}{M^2} (M^2 - 1) \right] \left( \alpha + 1 \right) - \alpha,
\label{eqn:peakT}
\end{equation}
in which $T_{i2}$ is the post-shock ion temperature, $\gamma$ is the adiabatic index for ions = 5/3, and $M$ is the pre-shock Mach number. In the original analysis of these data \cite{langendorf2018experimental}, the relative normal velocity of the two jets $v_{cs}$ was used as the input to eq.~(\ref{eqn:peakT}), resulting in the upper horizontal dashed lines as the predicted peak temperature in Figure~\ref{fig:dopplerTi}. Use of $v_{cs}$ overpredicts the expected collisional ion shock heating, as the relevant velocity to use is the normal velocity of one of the jets. Predicted ion shock heating in this case is shown as the lower dashed lines in Figure~\ref{fig:dopplerTi}.

Also plotted in Figure~\ref{fig:dopplerTi} are cooling rates calculated due to classical ion-electron collisional equilibration according to
\begin{equation}
\frac{dT_i}{dt} = \left[ \num{1.8e-19} \frac{ \left(m_i m_e \right)^{1/2}  \bar{Z}_i^2 n_e \Lambda_{ie} }
{ \left( m_i T_e + m_e T_i \right)^{3/2}} \right] \left( T_e - T_i \right),
\label{eqn:coolingTime}
\end{equation}
in which $m_i$ and $m_e$ are the ion and electron mass in grams, $T_e$ and $T_i$ are the electron and ion temperatures in eV, $\bar Z_i$ is the average ion charge state, $n_e$ is the electron number density in \si{cm^{-3}}, and $\Lambda_{ie}$ is the Coulomb logarithm for collisions between ions and electrons. These cooling trajectories are plotted as downward sloping dotted lines in Figure~\ref{fig:dopplerTi}. Also plotted in Figure~\ref{fig:dopplerTi} are the peak temperatures calculated in 1d multi-fluid CHICAGO simulations \cite{thoma2011two, thoma2017hybrid}, as filled stars. The filled stars are typically of similar value to the lower dashed line representing the predicted peak shock heating. Both of these results overall agree relatively well with measurements.

\begin{figure}[htbp]
\includegraphics[width=3.3truein]{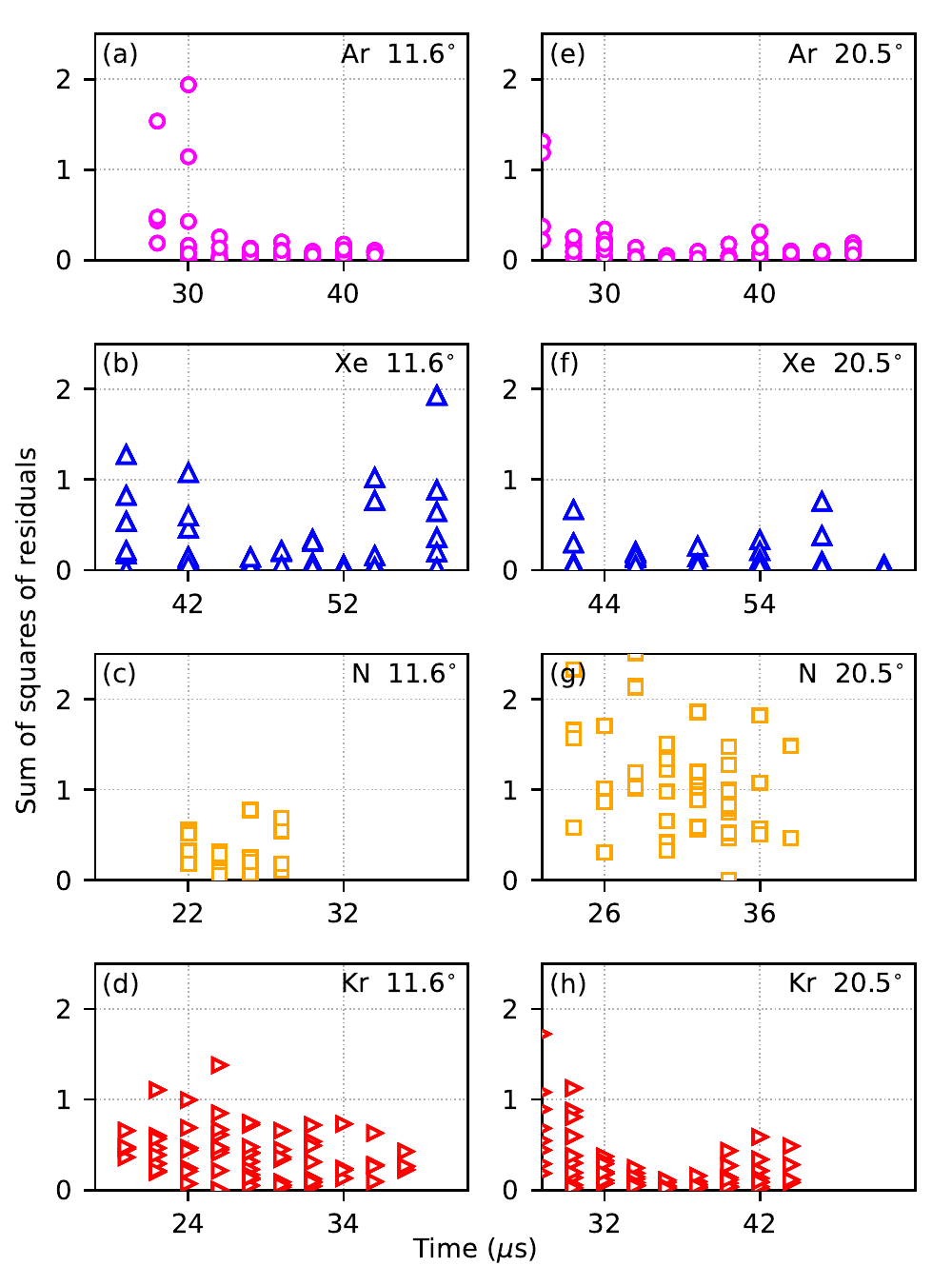}
\caption{Fitting residuals of ion temperature measurements. High values are often obtained at early times in the jet merger, consistent with the more varied line shapes observed (c.f. Fig \ref{fig:doppler} (a).}
\label{fig:dopplerResid}
\end{figure}

In the cases of argon and xenon at shallow merging angle, data points were accepted at early times and high temperatures much greater than the theoretical and simulation results. However a study of the fitting residuals (cf. Fig. \ref{fig:dopplerSpatial}, Fig. \ref{fig:dopplerResid}) shows that these points were accepted from experimental times with the high overall fitting residual, indicative of a relatively poor fit, so it is likely that these points are the remnants of the initial non-Gaussian signatures of the merging. 
In other cases, the simulation and theory results agree well with the data, with the data sometimes attaining slightly greater temperatures. Although this disagreement is often within experimental uncertainty, we speculate that it may be due to previously mentioned effects of the imperfect collimation of the jets leading to additional broadening in the observed line shape.

\section{Simulation Methodology}

We have used the plasma modeling code Chicago to perform 1-D
simulations of the transverse merging of jets in order to compare the
results to the measured ion temperatures obtained from the two-gun
experiments for Ar, N, Kr, and Xe colliding jets (both from guns at 
11.6\si{\degree} and 20.5\si{\degree} merging half angle). 
We have utilized the multi-fluid simulation capability of
Chicago to perform these simulations, in which Lagrangian
macroparticles convect fluid attributes (momentum, temperature,
internal energy, etc.)\cite{brackbill1988flip}.  The fluid momentum and
energy equations, along with Maxwell's equations are advanced on the
grid. This allows for solutions which are relatively free of numerical
diffusion and allows modeling with relatively few particles per
cell. Fluid particles may also be remapped to the Eulerian grid
\cite{thoma2011two}, which allows for the use of essentially a single
particle/cell/species, at the expense of a somewhat more numerically
diffusive solution. A direct-implicit electromagnetic field solve
\cite{welch2006integrated} allows for timesteps which are large compared to the
plasma period, and a cloud-in-cell particle weighting scheme
\cite{birdsall2004plasma} avoids numerical heating for cell sizes larger
than the Debye length. Equation of state information (charge state,
internal energy, pressure) is obtained from tables provided by the
PROPACEOS code \cite{macfarlane2006helios} \cite{macfarlane2007spect3d}. Radiation
transport is modeled by a multi-group diffusion treatment, with
opacities also obtained from PROPACEOS data, and is coupled to the
plasma through a source term in the electron energy equation.

\begin{table*}[tb]
\caption{Assessment of the ion heating and merging conditions to be anticipated in MJ-scale plasma liner experiments. It is observed that at increased densities, ion-electron equilibration is expected to be fast relative to the jet flight time, and that interpenetrating regimes may still be achievable at high liner energies.}
\label{tab:pjmif}
\begin{ruledtabular}
\begingroup
\setlength{\tabcolsep}{12pt} 
\begin{tabular}{ccccc}
Case&(1)&(2)&(3)&(4)\\
Case name&This work (a)&This work (e)&20 MJ slow&20 MJ fast\\
\hline
Jet speed (\si{km/s})&39&39&50&150\\
Chamber radius (\si{m})&1.36&1.36&4&4\\
Number of jets& & &600&600\\
Initial liner energy (\si{\mega\joule})& & &20&20\\
Length (radial) of merged liner (\si{m})& & &0.1&0.1\\
\hline
Liner merging radius $R_m$ (\si{m})& & &2.0&1.5\\
Liner avg. merge half-angle (\si{\degree})& & &4.7&4.7\\
Density at $R_m$ ($10^{15}$ \si{cm^{-3}})&0.23&0.18&15&2.8\\
\hline
$T_{i,pk}$ (eV)&6.7&18.3&4.9&38.9\\
$L_{ii,s}$ (\si{cm})&2.8&29.9&0.035&11.0\\
i-e equilibration time (\si{\micro\second})&19.7&23.9&0.6&2.8\\
Jet spot size at $R_m$ (\si{cm})&16&19&16&13\\
Flight time to chamber center (\si{\micro\second})&41.6&44.6&80&26.7\\
\hline
(I) Flight time $/$ ie equilibration time&2.1&1.9&130&9.6\\
(II) Jet spot size at $R_m$ $/~L_{ii,s}$&5.7&0.65&460&1.1\\
\end{tabular}
\endgroup
\end{ruledtabular}
\end{table*}%

\begin{figure}[htbp]
\includegraphics[width=3.34truein]{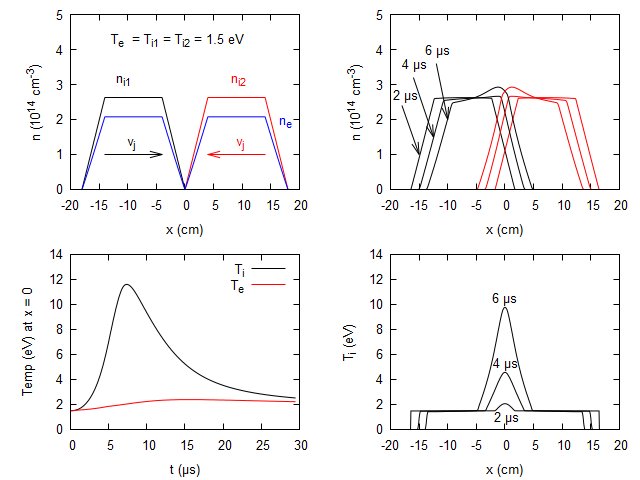}
\caption{Simulation setup and results for
1D Cartesian plasma jet merging simulation of argon jets.}
\label{fig:simsetup}
\end{figure}

\begin{figure}[htbp]
\includegraphics[width=3.34truein]{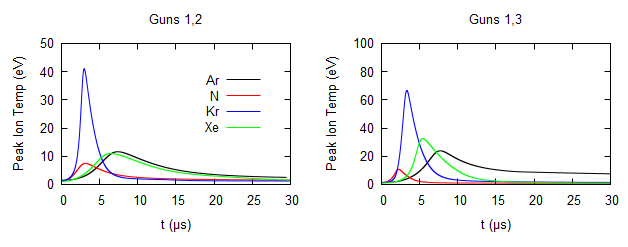}
\caption{Simulation peak ion temperatures as a function of time 
for runs corresponding to the experimental conditions.}
\label{fig:simpeakT}
\end{figure}

Figure~\ref{fig:simsetup} (top left) shows the simulation setup for a
1D Cartesian merging simulation for a pair of Ar jets, each with a
peak ion density of \num{2.63e14}~\si{cm^{-3}}, and a
(transverse) velocity of $8.36$~\si{km/s}. Each jet is treated as a
separate ion fluid species, which allows for interpenetration due to
finite ion-ion collisional mean free paths \cite{rambo1995comparison}. There is
also a single electron fluid species. The initial jet temperatures are
$1.5$~eV for the electron and both ion species. PIC implementation of
fluid modeling with a single electron and ion species is described in
some detail in Ref.~\cite{thoma2011two}, and the extension to allow for
multiple ion species is discussed in Ref.~\cite{thoma2017hybrid}. Radiation
transport is included in the single-group (gray) approximation. For
the density and temperature regimes in this simulation study it is
sufficient to assume a Spitzer collision frequency \cite{rambo1995comparison}
between charged-particle species. A cell-size of $0.06$~cm and
timestep of $1.3$~ps is used in all simulation results shown. With
these simulation parameters, we found negligible effects due to
numerical diffusion when using the Eulerian remapping procedure
described above, so it is used throughout. Global energy was tracked
during the simulations and found to be conserved to within a few
percent in most cases. In the top right of Fig.~\ref{fig:simsetup} we
have plotted the ion density profiles for each species at 2, 4, and
6~$\mu$s. Considerable jet interpenetration is visible, as is inferred in the experiments. In the
bottom right we have plotted the ``effective'' single ion temperature,
$T_i = (n_{i1} T_{i1} + n_{i2} T_{i2})/(n_{i1}+n_{i2})$, at the same
time increments. Fig.~\ref{fig:simpeakT} shows the peak ion
temperatures as a function of time.

\section{Implications for standoff plasma liner formation}
\label{sec:liner}

Using the results of this investigation, specifically the ion shock heating and equilibration results of section~\ref{sec:iontemp}, we can assess the previously raised concern of shock-heating-induced Mach number degradation of plasma liners formed by the merging of supersonic plasma jets. In the current experimental conditions, we do indeed observe substantial ion heating above the initial temperature persisting for many microseconds, which could be problematic if observed in a liner formation experiment. We can use the equations (\ref{eqn:peakT}) and (\ref{eqn:coolingTime}), supported by these experiments, to assess the likely heating for fusion-relevant liners.  We can estimate the merging radius $R_m$ \cite{langendorf2017semi}, local density $n_m$, merging half-angle $\theta$, and normal velocity $v_n$ as the following:
\begin{equation}
R_m = \left ( \frac{ r_{j0} (M + 1) + R_{w} } { 1 + \frac{2}{\sqrt{N}} \left( M+1 \right)} \right),
\label{eqn:mergingRadius}
\end{equation}
\begin{equation}
\cos(\theta) = 1 - \frac{2}{N},
\label{eqn:mergingHalfAngle}
\end{equation}
\begin{equation}
n_m = \frac{2E}{v^2}\left(\frac{1}{l_j(4 \pi R_m^2)}\right),
\label{eqn:mergingDensity}
\end{equation}
\begin{equation}
v_{n} = v_{jet} \sin(\theta),
\label{eqn:mergingRelVel}
\end{equation}
in which $r_{j0}$ is the initial jet radius, $R_w$ is the chamber wall radius, and $N$ is the number of discrete jets distributed around the sphere. With the density and relative velocity in hand, we can use equations (\ref{eqn:peakT}) and (\ref{eqn:coolingTime}) to assess the peak temperature attained and the time constant of subsequent cooling. Plugging in the velocity $v_n$ along with sound speed corresponding to a nominal pre-merge temperature, we obtain a peak ion temperature due to shock heating.  Taking this temperature, and density from (\ref{eqn:mergingDensity}), and plugging them in to equation (\ref{eqn:coolingTime}), we calculate the ion-electron relaxation time. Results are given in Table~\ref{tab:pjmif} for two cases with initial liner kinetic energy of 20 MJ, one with an implosion speed of 50 \si{km/s} and one with 150 \si{km/s}. In the slower case (3), we see that i-e equilibration is rapid in comparison to the liner flight time to the target (c.f. line (I) Table~\ref{tab:pjmif}), so it is clear that ion shock heating will not negatively impact the overall liner Mach number in this case. The interpenetration length is much smaller than the jet footprint on the merged liner (c.f. line (II) Table~\ref{tab:pjmif}), meaning that the merging should be collisional and result in shock formation.  

It is interesting to consider whether the jet merging could be accomplished in a similar collisionality regime to the 20.5\si{\degree} Argon case (e) studied in this work, in which interpenetration was observed to cause a relatively smooth merger. Considering the faster case (4) of Table~\ref{tab:pjmif} at 150 \si{km/s}, we see that this is indeed likely possible given the strong $v^4$ dependence of interpenetration distance on the jet relative velocity (c.f. line (II) Table~\ref{tab:pjmif}). In both MJ-scale cases, the number density is an order of magnitude higher than the current experiments. This should increase the rate of radiative cooling relative to electron heating via equilibration of shock-heated ions, as the former should scale as $n^2$ and the latter roughly as $n$. Therefore one would not expect an elevation in the plasma ionization state to cause a transition to collisional shock formation, as has been observed in e.g. Ref. \citenum{moser2015experimental}.

From this we first conclude that Mach number degradation due to ion shock heating should not be a significant effect to MJ-scale plasma liners at their currently envisioned velocities of order 50 \si{km/s}. We also note the intriguing possibility of a high-velocity jet merger operating in a somewhat interpenetrating regime, which may offer improved uniformity and symmetry of the merged plasma liner over the shock-forming case.

\section{Conclusions}
\label{sec:conclusions}

In conclusion, we observe nearly classical ion shock heating and ion-electron equilibration across a range of species, Mach numbers, and collisionalities.  In particular it is interesting to observe the differences and similarities between the jet mergers with interpenetration length of order of the mean free path, and in mergers with interpenetration length an order of magnitude higher. In such cases we see similar peak temperatures attained, but with an overall smoother merged structure with reduced density gradients. This structure is effectively a supersonic mutual collisional stopping similar to a collisional plasma shock, but occurring on the length scale of the ion interpenetration length and precluding the formation of abrupt density jumps on the order of the mean free path. In application to PJMIF liners, we find that Mach number degradation due to ion shock heating will likely not be significant at the typical full-scale conditions proposed. In addition, the smoothness of the jet merger and decrease in density gradients observed in interpenetrating cases may make them an attractive candidate for approaches such as PJMIF which seek to form smooth and uniform structures from the merging of discrete supersonic plasma sources.


%
%

%

\begin{acknowledgments}
We acknowledge J. Dunn, E. Cruz, A. Case, F.D. Witherspoon, S. Brockington, J. Cassibry, R. Samulyak, P. Stoltz, Y. C. F. Thio, and D. Welch for technical support and/or useful discussions. This work was supported by the Advanced Research Projects Agency -- Energy and the Office of Fusion Energy Sciences of the U.S. Dept. of Energy under Contract No. DE-AC52-06NA25396.
\end{acknowledgments}

\bibliography{pop_shock_ion}

\end{document}